\documentclass[12pt,a4paper,final]{article}

\usepackage[backref,
            natbib
            ,style = numeric-comp
            ,maxnames = 2
            ,backend = bibtex
            ,sorting=none
            ]{biblatex}
  
\usepackage[T1]{fontenc} 
\usepackage{palatino,eulervm}
\usepackage[english]{babel} 
\usepackage{csquotes}
\usepackage{amsmath,amssymb,amsfonts,amsthm} 
\usepackage[table,dvipsnames*,svgnames]{xcolor}
\usepackage{fixltx2e}
\usepackage{showkeys}

\usepackage[text={490pt,650pt},centering]{geometry} 
\usepackage[active]{srcltx}
\usepackage{extarrows}

\usepackage{braket}
\usepackage{mathrsfs}

\makeatletter

\let\@authors\@empty
\let\@email\@empty
\let\@affiliation\@empty
\let\@pdfsubject\@empty
\let\@keywords\@empty
\let\@preprint\@empty

\providecommand{\pdfsubject}[1]{\gdef\@pdfsubject{#1}}
\providecommand{\keywords}[1]{\gdef\@keywords{#1}}
\renewcommand{\author}[1]{\ifx\@authors\@empty\toks@\expandafter{#1}\else\toks@\expandafter{\@authors, #1}\fi\edef\@authors{\the\toks@}}
\providecommand{\email}[1]{\ifx\@email\@empty\toks@\expandafter{#1}\else\toks@\expandafter{\@email, #1}\fi\edef\@email{\the\toks@}}
\providecommand{\affiliation}[1]{\gdef\@affiliation{#1}}
\providecommand{\preprint}[1]{\gdef\@preprint{#1}}

\makeatother

\usepackage[hyperindex,breaklinks]{hyperref}

\hypersetup{
    unicode=false,          
    pdftoolbar=true,        
    pdfmenubar=true,        
    pdffitwindow=false,     
    pdfstartview={FitH},    
    pdfproducer={Brown University}, 
    pdfnewwindow=true,      
    colorlinks=true,       
    linkcolor=black,          
    citecolor=DarkGreen,        
    filecolor=magenta,      
    urlcolor=cyan!30!black!70         
}

\usepackage{graphicx}

\allowdisplaybreaks[3]
\newcommand{\del}{\partial}
\newcommand{\ddel}[2]{\frac{\partial #1}{\partial #2}}

\newcommand{\bal}{\begin{align}}
\newcommand{\eal}{\end{align}}
\newcommand{\scri}{\mathscr{I}}

\renewcommand{\[}{\begin{equation}}
\renewcommand{\]}{\end{equation}}

\newcommand{\R}{\mathbb{R}}

\bibliography{BMS.bib}

\begin{document}

\title{BMS, String Theory, and Soft Theorems}

\author{Steven G.\ Avery}
 \email{\href{mailto:steven\_avery@brown.edu}{steven\_avery@brown.edu}}

\author{Burkhard U.\ W.\ Schwab}
 \email{\href{mailto:burkhard\_schwab@brown.edu}{burkhard\_schwab@brown.edu}}

\affiliation{%
Brown University\\Department of Physics\\182 Hope St, Providence, RI, 02912}

\keywords{String Theory, BMS}
\pdfsubject{Soft Gravitons}
\preprint{Brown-HET-1674}


\makeatletter
\thispagestyle{empty}

\begin{flushright}
\begingroup\ttfamily\@preprint\par\endgroup
\end{flushright}

\begin{centering}
\begingroup\Large\normalfont\bfseries\@title\par\endgroup
\vspace{1cm}

\begingroup\@authors\par\endgroup
\vspace{5mm}

\begingroup\itshape\@affiliation\par\endgroup
\vspace{3mm}

\begingroup\ttfamily\@email\par\endgroup
\vspace{0.25cm}

\begin{minipage}{17cm}
 \begin{abstract}
We study the action of the BMS group in critical, bosonic string theory living on a target space of the form $\mathbb{M}^{d}\times C$. Here $M^{d}$ is $d$-dimensional (asymptotically) flat spacetime and $C$ is an arbitrary compactification. We provide a treatment of generalized Ward--Takahashi identities and derive consistent boundary conditions for any $d$ from string theory considerations. Finally, we derive BMS transformations in higher dimensional spacetimes and show that the generalized Ward--Takahashi identity of BMS produces Weinberg's soft theorem in string theory.
 \end{abstract}
\end{minipage}

\vspace{3mm}
\rule{\textwidth}{.5mm}
\vspace{-1cm}

\end{centering}

\makeatother

\tableofcontents

\section{Introduction}
\label{sec:introduction}

It has long been known that asymptotically flat spacetimes in three and four dimensions allow an
enhanced group of large diffeomorphisms~\cite{Bondi:1962px,Sachs:1962wk}. In these cases, the global
Poincar\'e symmetry gets enlarged to the so called Bondi--van der Burg--Metzner--Sachs (BvBMS, but
most often just BMS) symmetry. The resulting symmetry extends the translations into an infinite
dimensional Abelian group known, historically and somewhat unfortunately, as supertranslations:
angle-dependent translations along $\mathscr{I}^{\pm}$. These large diffeomorphisms are not
isometries of flat space, but rather take one asymptotically flat solution into another.

Another famous example of enhanced asymptotic symmetry is gravity in asymptotically
AdS\textsubscript{3} spacetime~\cite{Brown:1986nw}.  The SL(2,$\mathbb{C}$) isometry of global
AdS\textsubscript{3} becomes a subalgebra of the asymptotic Virasoro symmetry in accord with
conformal symmetry in two dimensions, as is required by AdS\textsubscript{3}/CFT\textsubscript{2}
duality~\cite{Aharony:1999ti}. Just as a general Virasoro transformation is not a symmetry of the
CFT\textsubscript{2} vacuum, the large diffeomorphisms in~\cite{Brown:1986nw} are not isometries of
AdS\textsubscript{3}. AdS/CFT duality implies that the enhanced symmetry can be realized on the
string worldsheet theory in asymptotically AdS\textsubscript{3} space. Indeed, the generators of
target space Virasoro transformations were found and the algebra
studied in~\cite{Giveon:1998ns,Kutasov:1999xu,deBoer:1998pp, Giveon:2001up,
  Troost:2010zz,Ashok:2009jw}.  The resulting Virasoro algebra is a subset of the algebra of target
space diffeomorphisms and may be understood as the set of diffeomorphisms that are neither pure
gauge degrees of freedom nor global symmetries of the theory. Their action is to insert physical
degrees of freedom which should be gravitons; however, since there are no propagating gravitons in
three dimensions, these degrees of freedom live on the boundary of AdS\textsubscript{3}.

Compared to AdS spacetimes, on asymptotically flat space, large diffeomorphisms are a much less well
developed topic. This has various reasons, one being the much harder boundary structure of
asymptotically flat space times. Here the boundary is composed of past and future timelike infinity
$i^{\pm}$, spacelike infinity $i^{0}$, and past and future null infinity $\mathscr{I}^{\pm}$. In a
recent development, the fate of massless particles on such spacetimes, most notably gravitons and
photons in four dimensions has been
addressed~\cite{Strominger:2013lka,Strominger:2013jfa,Kapec:2014zla,Kapec:2014opa,He:2014laa,He:2014cra}. These
particles start on $\mathscr{I}^{-}$ and end on $\mathscr{I}^{+}$.

In the semiclassical approach to quantum gravity used in~\cite{Strominger:2013jfa}, Strominger
argued that the diagonal subgroup
BMS\textsuperscript{0}$\subset$BMS\textsuperscript{+}$\times$BMS\textsuperscript{-} is a
spontaneously broken symmetry of the quantum gravity $\mathcal{S}$-matrix. It was shown that the
degrees of freedom associated with supertranslations can be understood as soft, i.e. zero momentum,
gravitons. This is similar to ideas in AdS\textsubscript{3}, however, in asymptotically flat $d=4$
spacetime the physics is richer: this statement directly connects the large diffeomorphisms of
four-dimensional asymptotically flat spacetime with Weinberg's soft theorem ``in the bulk''. The
interest in the soft behavior of gravitons and, by extension, gluons and photons, has seen a
renaissance due to these developments. Let us quickly mention a (non-exhaustive) list of recent
work.

The subleading corrections to Weinberg's soft theorem have been investigated in various theories in
\cite{Schwab:2014xua,Zlotnikov:2014sva,Kalousios:2014uva,Schwab:2014fia,Schwab:2014sla,Bianchi:2015yta,Bianchi:2014gla,DiVecchia:2015oba,Bern:2014oka,Bianchi:2014gla,He:2014bga,Volovich:2015yoa}. In
the case of ambitwistor string theory, the appearance of the soft limits is quite well understood
already \cite{Geyer:2014lca,Adamo2014,Adamo:2015fwa,Lipstein:2015rxa}. Some results for the case of
gauge theories may be found in
\cite{Casali:2014xpa,Broedel:2014bza,Broedel:2014fsa,Bonocore:2014wua,White:2014qia}.

In higher dimensional spacetimes, such a symmetry enhancement has been ruled out historically. This
conclusion has been based on the use of the most stringent boundary conditions that still allow for
gravitational radiation to reach asymptotic infinity. However, Weinberg's soft theorem is unaltered
in higher dimensions. This suggests the use of more lenient boundary conditions that allow for BMS
type symmetries in $D>4$. Such a solution has been suggested in~\cite{Kapec:2015vwa} and this is
what we advocate in this paper in the context of string theory. Interestingly, it seems that string
theory even \emph{prefers} boundary conditions that allow for the discussed symmetry enhancement in
any dimension.

We discuss how string theory realizes large diffeomorphisms in target space as quasi-symmetries from
the vantage point of the path integral. In the semi-classical treatment, BMS transformations have been
shown to be spontaneously broken symmetries and the soft theorem to be a Ward--Takahashi identity. In the
worldsheet theory of string theory this is of course not the correct mechanism. Target space
diffeomorphisms are field redefinitions in the worldsheet theory. These are, apart from the Poincar\'e symmetry, not actual symmetries of the worldsheet theory. In fact, the only reason
why target space diffeomorphisms are allowed is because they either insert unphysical graviton
states into the theory, which have no effect on observables like the $\mathcal{S}$-matrix, or they
insert physical gravitons into the theory. (In flat space these are soft gravitons with zero
momentum, whereas in AdS\textsubscript{3} these are ``boundary gravitons''.) For this reason, these
transformations belong in the category of dualities between string models and should not be confused
with gauge symmetries. Additionally, we give a detailed study of the difference between global
symmetries, large diffeomorphisms, and small diffeomorphisms. Recently, there has been some
investigation along these lines~\cite{Schulgin:2013xya,Schulgin:2014gra}, although with a different
emphasis and perspective.

The paper is organized as follows. In sec.~\ref{sec:ward} we discuss Ward--Takahashi identities in
the case of spontaneously broken symmetries and more importantly, general transformations of the
action. In the following sec.~\ref{sec:gravity} we discuss various topics related to asymptotic
flatness, gauge choices in gravity and BMS symmetries. In this chapter we also try to clear up a few
points of confusion in the literature. Finally, in sec.~\ref{sec:stringasympspc} we discuss the
measure of the string theory path integral in curved backgrounds, the difference between large,
small and forbidden diffeomorphisms from the worldsheet perspective, as well as boundary conditions
in $d$ dimensions. The chapter finishes with a discussion of Weinberg's soft theorem as a
generalized Ward--Takahashi identity on the worldsheet. In the conclusions, we speculate on some
topics related to the subleading soft theorem in higher dimensions, the apparent preference of
string theory for specific boundary conditions, and some other topics.

\section{Ward--Takahashi Identities}
\label{sec:ward}

To make our argument, we need two generalizations of the usual Ward--Takahashi (WT) identities. To
that end, let us first recall the standard argument. For concreteness, we follow the approach
in~\cite{Polchinski:1998rq}.

Consider a field theory with action functional $S[\phi]$ that is invariant under an infinitesimal
field transformation $\delta\phi$:
\begin{equation}
S[\phi + \delta \phi] = S[\phi].
\end{equation}
Classically, this implies a locally conserved current. In quantum field theory, one can show that
the current is conserved up to contact terms (when inserted into a correlator). Consider a path
integral with local insertions
\begin{equation}\label{eq:gen-correlator}
\braket{\Phi_1(x_1)\cdots \Phi_n(x_n)}  = \int D\phi \big(\Phi_1(x_1)\cdots \Phi_n(x_n)\big)e^{i S[\phi]},
\end{equation}
and perform the change of variables
\begin{equation}
\phi(x) \mapsto \phi(x) + \rho(x)\delta\phi(x),
\end{equation}
with $\rho(x)$ an arbitrary function with compact support. Since $\rho = \text{const.}$ is a
symmetry and since this is a local transformation, we must have
\begin{equation}\label{eq:jdelrho}
\delta_\rho S = \int d^d x\, j^\mu \del_\mu \rho.
\end{equation}
This is just a change of variables; we don't change the value of the left-hand side
of~\eqref{eq:gen-correlator}, and thus, assuming the measure is invariant,
\begin{equation}
\int D\phi \left[\delta_\rho\big(\Phi_1(x_1)\cdots \Phi_n(x_n)\big) + i\big(\Phi_1(x_1)\cdots \Phi_n(x_n)\big) \int d^d x\, j^\mu \del_\mu \rho\right] e^{i S[\phi]} = 0.
\end{equation}
This must hold for general $\rho(x)$, and therefore the local WT identity 
\begin{equation}\label{eq:local-WT}
\braket{\Phi_1(x_1)\cdots \Phi_n(x_n) \del_\mu j^\mu(y)} 
  = -i \sum_j \delta^{(d)}(x_j-y) \braket{\Phi_1(x_1) \cdots \delta_j\Phi(x_j)\cdots \Phi_n(x_n)} 
   + \ddel{F^\mu}{y^\mu}
\end{equation}
holds for some undetermined $F$, which we usually omit. The current $j$ is the usual locally conserved
Noether current from classical mechanics.

There is a weaker global version of the above WT identity, which we find useful for our discussion
of BMS. Just as classically the locally conserved current implies the existence of a globally
conserved charge $Q$, we can ask for the analogous statement in quantum field theory. For this
purpose, it is simplest to switch to an operator language.

The charge $Q$ generates the symmetry transformation of the local operators,
\begin{equation}
[Q,\,\Phi(x)] = \delta \Phi(x).
\end{equation}
Now consider the correlator
\begin{equation}
\Braket{\big[Q,\, \Phi_1(x_1)\cdots \Phi_n(x_n)\big]} = \bra{0} T \big[Q,\, \Phi_1(x_1)\cdots \Phi_n(x_n)\big] \ket{0}.
\end{equation}
Assuming the vacuum is invariant under the transformation, $Q\ket{0} = 0$, then one finds the global
WT identity
\begin{equation}\label{eq:global-WT}
\Braket{\big[Q,\, \Phi_1(x_1)\cdots \Phi_n(x_n)\big]} = 0.
\end{equation}
To wit, the only non-vanishing correlators must be singlets of the symmetry transformation.

\subsection{Spontaneously Broken Symmetry}
\label{sec:ward-SSB}

In the presence of spontaneous symmetry breaking, the WT identities are modified~\cite{Matsumoto:1973hg,Matsumoto:1974nt}. For the
local WT identity~\eqref{eq:local-WT}, the total divergence term $\del_\mu F^\mu$ cannot be dropped and
is fixed by the transformation of the vacuum. To derive the local WT identities, one adds a small
parameter $\epsilon$ times a symmetry breaking term in the path integral that selects the correct
vacuum. In the end, one then carefully takes $\epsilon$ to zero to find the nontrivial
contributions. We refer the interested reader to the references~\cite{Matsumoto:1973hg,Matsumoto:1974nt,Nakanishi:1974pz} for more details.

For our purposes, it suffices to consider the fate of the global WT identity~\eqref{eq:global-WT}
under spontaneously broken symmetry. (Here, we refer to the treatment of~\cite{Nakanishi:1974pz}.) Assuming the
symmetry generator $Q$ commutes with the Poincar\'e generator $P_\mu$, then $Q\ket{0}$ must give a new
vacuum $\ket{0'}$. Since the different vacua are direct-producted into the rest of the Hilbert
space, this suggests a decomposition of the symmetry generator $Q$ as
\begin{equation}
Q = Q_\text{soft} + Q_\text{hard},
\end{equation}
where $Q_\text{soft}$ acts only on the vacuum and $Q_\text{hard}$ transforms the nonzero energy
states and annihilates the vacuum. Formally, then, the global WT identity takes the form
\begin{multline}\label{eq:global-WT-SSB}
\bra{0}T\big[Q_\text{hard},\, \Phi_1(x_1)\cdots \Phi_n(x_n)\big]\ket{0} + \big(\bra{0}Q_\text{soft}\big)T \Phi_1(x_1)\cdots \Phi_n(x_n)\ket{0} \\
  - \bra{0}T\Phi_1(x_1)\cdots \Phi_n(x_n)\big(Q_\text{soft}\ket{0}\big) = 0,
\end{multline}
where we assume the insertions $\Phi_i$ are all hard. How does $Q_\text{soft}$ act on the vacuum?
Assuming it is a linear function of the fields, then it must be proportional to inserting a single
Goldstone boson. The WT identity then relates a correlator with soft insertions to the correlator
with transformed (hard) insertions. We have written this in a form that is suggestive of the
argument for the soft graviton theorem from supertranslation
symmetry~\cite{Strominger:2013jfa,He:2014laa}, a point we return to in Section~\ref{sec:softSec}.

\subsection{General Transformations}

The above generalization of WT identities is appropriate when considering BMS as a symmetry of the
$\mathcal{S}$-matrix in a spacetime description. In the worldsheet description, however, BMS transformations
and, indeed, all target space diffeomorphisms are \emph{not} Noetherian symmetries of the worldsheet
action. Thus, we must consider a different generalization of WT identities: not for spontaneously
broken symmetry, but for general transformations.\footnote{These might also be referred to as
  Schwinger--Dyson equations; however, in the application of interest the identities are related to
  a more general notion of symmetry (duality) and so we will think of them as generalized WT identities.}

Consider a local, infinitesimal transformation of the fields $\delta \phi(x)$. Define $V(x)$ in
terms of the transformation of the action
\begin{equation}
\delta S[\phi] = \int d^d x\, V(x).
\end{equation}
As we shall see, $V(x)$ acts a source, giving the classical violation of current conservation.  As
in the original derivation of the WT identities, let us now insert a local function $\rho(x)$, so
that
\begin{equation}
\delta_\rho \phi(x) = \rho(x)\delta\phi(x).
\end{equation}
Since the variation of the action is the integral of $V$ when $\rho$ is constant, we have
\begin{equation}
\delta_\rho S[\phi] = \int d^d x\left[\rho(x) V(x) + j^\mu\del_\mu \rho\right].
\end{equation}
Proceeding as above, we now find an extra term in the WT identity:
\begin{equation}
\int D\phi \left[\delta_\rho\big(\Phi_1(x_1)\cdots \Phi_n(x_n)\big) 
+ i\big(\Phi_1(x_1)\cdots \Phi_n(x_n)\big) \int d^d x\, \left( \rho(x) V(x) + j^\mu \del_\mu \rho\right)\right] e^{i S[\phi]} = 0.
\end{equation}
Varying $\rho$ and integrating by parts, we find the generalization of~\eqref{eq:local-WT}
\begin{multline}
\braket{\Phi_1(x_1)\cdots \Phi_n(x_n) \del_\mu j^\mu(y)} = \braket{\Phi_1(x_1)\cdots \Phi_n(x_n) V(y)}\\
  -i \sum_j \delta^{(d)}(x_j-y) \braket{\Phi_1(x_1) \cdots \delta\Phi_j(x_j)\cdots \Phi_n(x_n)} ,\label{eq:anomWard}
\end{multline}
where we have dropped the total divergence term, since we do not need it. When we apply this to
target space diffeomorphisms of the worldsheet theory, we find that the vertex operator term $V$ gets
related to the extra term in~\eqref{eq:global-WT-SSB} for the spacetime theory. 

\section{Asymptotically Flat Spacetimes and BMS Transformations}
\label{sec:gravity}

Let us review the relationship between BMS symmetry and Weinberg's soft theorem, as developed
in~\cite{Strominger:2013jfa,Kapec:2014opa,He:2014laa}. This is the target space story; we are
interested in reproducing and reinterpreting these results by studying the string worldsheet theory.

The original BMS symmetry~\cite{Bondi:1962px,Sachs:1962wk} was in four dimensions; however, many
authors have considered extensions in higher dimensions~\cite{Barnich:2013sxa, Banks:2014iha,
  Hollands:2003ie, Hollands:2003xp, Ishibashi:2007kb, Tanabe:2009va, Tanabe:2011es,
  Hollands:2004ac}. In light of the recent argument for Weinberg's soft graviton theorem in even
dimensions from BMS~\cite{Kapec:2014opa}, we will hold our discussion for general $d$-dimensional
spacetimes.

\subsection{Bondi Gauge}

Consider an asymptotically flat spacetime. There are many notions of asymptotic flatness in the
literature; for our purposes, let us define an asymptotically flat metric as one that in a
neighborhood of $\scri^+$ (and similarly for $\scri^-$) may be written in $d$-dimensional Bondi
gauge,
\begin{equation}\label{eq:bondi-gauge}
g_{r r} = g_{rA} = 0\qquad 
\partial_r \det \left(\frac{g_{AB}}{r^2}\right)= 0,
\end{equation}
in the form
\begin{equation}\label{eq:flat-bondi}
ds^2 = -du^2 - 2du dr + r^2\gamma_{AB}d\theta^Ad\theta^B + \left(\text{subleading in }\tfrac{1}{r}\right),
\end{equation}
where $u$ is a null coordinate along $\scri$, $r$ a radial coordinate, and $\{\theta^A\}$
coordinates on the unit sphere $S^{d-2}$ with metric $\gamma_{AB}$. When needed, we use the
following metric for the unit $(d-2)$-sphere:
\begin{equation}
ds_{S^{(d-2)}}^2 = \frac{4}{(1+\vec{\theta}^2)^2}d\vec{\theta}^2\qquad \vec{\theta}\in \R^{d-2},
\end{equation}
where $\vec{\theta}^2$ denotes the usual Cartesian product.  

For general $d$, the fall-off conditions one imposes on the subleading terms has been the subject of
some discussion~\cite{Hollands:2003ie, Hollands:2003xp, Ishibashi:2007kb, Tanabe:2009va,
  Tanabe:2011es, Hollands:2004ac}. The traditional point of view, at least as far as the authors of
this document understand, is that one should impose the \emph{most} restrictive boundary conditions
that still allow gravitational radiation to hit $\scri$. For $d>4$, these restrictive boundary
conditions eliminate the enhanced symmetry that occurs in four dimensions, since the radiative modes
fall off faster in higher dimensions.

An alternative philosophy might be to impose the \emph{least} restrictive boundary conditions
consistent with~\eqref{eq:flat-bondi} and the equations of motion. This approach is essentially what
is considered in~\cite{Kapec:2015vwa}, and what allows a nontrivial BMS algebra in higher
dimensions. Let us note that as long as the boundary conditions are consistent, there is no a priori
``correct'' boundary condition: different boundary conditions define different semiclassical
gravitational theories. This is an idea that should be familiar from
AdS/CFT.\footnote{See~\cite{Balasubramanian:1998sn, Klebanov:1999tb} for an early explicit instance,
  and~\cite{Compere:2008us, Poojary:2014ifa, Apolo:2015fja, Apolo:2014tua, Avery:2013dja,
    Compere:2013bya,Troessaert:2013fma} for some recent, more exotic examples.}

Specifically, the ``traditional'' fall-off conditions are where the radiative data enter
asymptotically~\cite{Tanabe:2011es}:
\begin{equation}\begin{aligned}\label{eq:radiative-BCs}
g_{uu} &= -1 + O\left(r^{-\frac{d-2}{2}}\right)\\
g_{ur} &= -1 + O\left(r^{-(d-2)}\right)\\
g_{uA} &= O\left(r^{-\frac{d-4}{2}}\right)\\
g_{AB} &=r^2\gamma_{AB} + O\left(r^{\frac{6-d}{2}}\right).
\end{aligned}\end{equation}
On the other
hand, the weaker boundary conditions employed in~\cite{Kapec:2015vwa} are the same as in four
dimensions:
\begin{equation}\begin{aligned}\label{eq:asymp-flat}
g_{uu} &= -1 + O\left(r^{-1}\right)\\
g_{ur} &= -1 + O\left(r^{-2}\right)\\
g_{uA} &= O\left(r^{0}\right)\\
g_{AB} &=r^2\gamma_{AB} + O\left(r^{1}\right),
\end{aligned}\end{equation}
with additional dimension-dependent conditions on the Ricci tensor (and the stress-tensor in
theories with a matter sector). While the additional constraints are likely important to have a
consistent theory, we do not need them in our discussion.

\subsection{Large versus Small Diffeomorphisms}

Gravity can be understood as a gauge theory of diffeomorphism symmetry. In a putative quantization
of gravity, one expects to identify states that differ only by a diffeomorphism; however, there are
a class of diffeomorphisms that fall off slowly enough at large $r$ to affect the radiative data and
give finite asymptotically conserved charges. These ``large diffeomorphisms''\footnote{Note that this is a
  different usage of the phrase ``large diffeomorphism'' than when it is used to mean diffeomorphisms that
  are not smoothly connected to the identity.} should not be modded out of the Hilbert space, in
contradistinction to the familiar ``small diffeomorphisms''. In gauge theory, there is an analogous
statement, that one does not mod out by the global part of the gauge group. This issue only arises
because we have a manifold with boundary, and boundary conditions. This means that there can also be
forbidden diffeomorphisms that violate the boundary conditions. The asymptotic symmetry group can be thought
of as the small diffeomorphism equivalence classes of the allowed diffeomorphisms.

The classification of large versus small diffeomorphisms is part of the definition of the theory that
depends on the boundary conditions.

\subsection{BMS}
\label{sec:bms-algebra}

What is the algebra of large diffeomorphisms for asymptotically flat space? Naively, one would expect to
recover the global part of the gauge group, the Poincar\'e group. In four dimensions, for any boundary
conditions that allow gravitational waves at $\scri$, the Poincar\'e symmetry is enhanced to the
infinite-dimensional BMS\textsubscript{4} algebra~\cite{Bondi:1962px,Sachs:1962wk}. For $d>4$, there
is enhancement to BMS\textsubscript{$d$} if one uses the weaker boundary
conditions~\eqref{eq:asymp-flat}.

One can find the algebra of large diffeomorphisms, by solving for the set of infinitesimal
diffeomorphisms that preserve the form~\eqref{eq:asymp-flat}. That is, vector fields whose Lie
derivative of a metric satisfying~\eqref{eq:asymp-flat} give variations at the same order as
in~\eqref{eq:asymp-flat}. (One may also follow the covariant treatment in~\cite{Barnich:2001jy} that precisely
separates out the equivalence classes.) The vector fields
\[
\xi = \xi^{u}\del_{u} + \xi^{r}\del_{r} + \xi^{A}\del_A,
\]
generating the BMS group take the form
\begin{equation}\begin{aligned}\label{eq:BMSgen}
\xi^{u} &= T(\theta) + u\, g(\theta) \\
\xi^{r} &= -r \,g(\theta) + \frac{1}{d-2}\Delta (T(\theta) + u\, g(\theta)) + \dots\\
\xi^{A} &= Y^A(\theta) - \frac{1}{r}D^A\left(T(\theta) + u\, g(\theta)\right) + \dots,
\end{aligned}\end{equation}
where $T(\theta)$ is a general function on the sphere, and $Y^A$ is a conformal Killing vector (CKV) of
$S^{d-2}$ with conformal factor $g$:
\begin{equation}
\label{eq:confKilvec}
\mathcal{L}_Y\gamma_{AB} = 2g(\theta)\gamma_{AB},\qquad g(\theta) = \frac{1}{d-2}D_AY^A(\theta).
\end{equation}
The omitted subleading terms in~\eqref{eq:BMSgen} are ``small'' metric-dependent terms that do not
contribute to asymptotic charges or alter the algebra. $\Delta$ is the Laplacian with respect to the round metric
$\gamma$ and $D_{A}$ is its covariant derivative. We raise its index with $\gamma^{AB}$.

The sphere $S^{d-2}$ is maximally symmetric, and therefore admits a full set of $\frac{d(d-1)}{2}$
linearly independent CKVs, $Y^A$. With $T=0$, these generate the $\frac{d(d-1)}{2}$ Lorentz
transformations of $SO(d-1,1)$. If one uses the more restrictive
conditions~\eqref{eq:radiative-BCs}, then for $d>4$ one finds the additional requirement that $D^A T$ satisfy
the sphere's conformal Killing equation. For a scalar function, one can show that there exist $d$
linearly independent solutions, which generate the $d$ ordinary translations of the Poincar\'e
group. The vector fields~\eqref{eq:BMSgen} with unconstrained $T(\theta)$ and $Y^A=0$ generate the
supertranslations.

Thus one sees that the BMS\textsubscript{$d$} group may be written as the semidirect product,
\[
\textrm{BMS}_d = T\rtimes SO(d-1,1),
\] 
non-singular transformations of the sphere at null infinity $\scri$---the rotations---acting on the
infinite-dimensional Abelian group $T$ of supertranslations.  In four dimensions, the infinitesimal
version of the first group can be augmented to a Virasoro algebra
\cite{Barnich:2011ct,Barnich:2011mi} called superrotations. The resulting semidirect sum of
superrotations acting on supertranslations is known as the \emph{extended} BMS\textsubscript{$4$}
algebra. In the conclusion, we speculate on
the extension of $\textrm{BMS}_d$ for general $d$, a nontrivial problem since the conformal group of
the sphere is only enhanced for $S^2$.

\subsection{The Soft Theorem, Part I: The Gravitational S-Matrix}
\label{sec:spacetime-soft-theorem}

One may study the asymptotic geometry at $\scri^+$ and $\scri^-$, and find two sets of apparently
disconnected transformations, which following~\cite{Strominger:2013jfa} we denote BMS\textsuperscript{$+$} and
BMS\textsuperscript{$-$}. In the case of flat space, the Bondi coordinates appropriate to both
$\scri^+$ and $\scri^-$ cover the entire space, with an easily established coordinate
transformation. One may then observe that the usual Poincar\'e subalgebra generators of $\scri^+$ are
the same as those of $\scri^-$, when one makes an antipodal identification of the sphere. This
should not be surprising since light rays starting from a point on $\scri^-$ travel to the antipodal
point on $\scri^+$.

Thus, following~\cite{Strominger:2013jfa}, let's define the diagonal BMS group,
BMS\textsuperscript{0} via this antipodal map. We may then conjecture\footnote{If one restricts
  oneself to Christodoulou--Klainerman spaces, as does~\cite{Strominger:2013jfa}, then one can \emph{show}
  without conjecture, but this presentation of the argument seems cleaner and more general.} that
the Poincar\'e symmetry of the gravitational $\mathcal{S}$-matrix is enhanced to BMS\textsuperscript{0}
symmetry. Indeed, depending on how seriously one takes BMS symmetry, one might even wish to demand
that this hold for any consistent quantization of gravity in asymptotic flat space. Since
supertranslations are not isometries of flat space, the supertranslation symmetry must be
spontaneously broken. Thus, if we wish to write a WT identity for supertranslations, we are in the
purview of the discussion in sec.~\ref{sec:ward-SSB}.

Before proceeding, it is amusing to ask: what field gets a vacuum expectation value, signaling the
spontaneous breaking of supertranslation symmetry? The answer is apparently the metric itself. While
the metric is not gauge invariant, the asymptotic form \emph{is} as is clear from our discussion of
large diffeomorphisms. The symmetry restoration phase, then, is when there is no spacetime (or at
least no way to measure distances) at all, $g_{\mu\nu} \equiv 0$, whatever that means. If we want to
find the local WT identity from the path integral following~\cite{Matsumoto:1974nt}, then we expect
an appropriate boundary term in the action to play the role of the symmetry breaking term. We leave
these speculations to be developed more fully elsewhere.

Proceeding to the soft theorem, let us first note that supertranslations commute with $\partial_u$
(and the other translation generators), and therefore do not change the energy (or momentum): if
$Q_T$ is the generator of a supertranslation (in BMS\textsuperscript{0}) parametrized by
$T(\theta)$, then $Q_T$ acting on the vacuum must give a new vacuum state as in the discussion
of~\ref{sec:ward-SSB}. This is an important distinction from Virasoro symmetry in
AdS\textsubscript{3}, where the Virasoro generators $L_{-n}$ that are not isometries of
AdS\textsubscript{3} take the global AdS vacuum into \emph{excited states}; they are not soft
transformations.

Following~\cite{Strominger:2013jfa}, we introduce a basis of asymptotic states via Fock space modes
with energy labelled by $\partial_u$ eigenvalue, $E$, and direction labelled by points on $S^{d-2}$,
$\theta$. Since we have argued the vacuum is degenerate, it should also carry a label telling us
which vacuum state we are in, but we will suppress this.  This should correspond to all possible
occupation numbers of soft graviton modes. Let's just call this data collectively $\psi$ for now;
thus, we write the vacuum as $\ket{0;\psi}$. The zero mode label $\psi$ is direct producted into the
hard Fock states.  Thus, for instance a one particle state in this language takes the form $\ket{E,
  \theta; \psi}$, suppressing polarization data or other internal quantum numbers of the
excitation. Formally, then, we may decompose $Q_T\in \text{BMS}^0$ into the sum $Q_T^\text{soft} +
Q_T^\text{hard}$ such that
\begin{equation}
Q_T^\text{hard}\ket{0;\psi} = 0.
\end{equation}
The hard part of the charge, $Q_T^\text{hard}$, just performs the supertranslation on the Fock
states with nonzero energy. For instance, on a single (hard) particle state, we have
\begin{equation}
Q_T^\text{hard}\ket{E,\theta; \psi} = T(\theta)\,E \ket{E,\theta;\psi},
\end{equation}
since the large part of the supertranslation is $T(\theta)\partial_u$. $Q_T^\text{soft}$ gives the
action of a supertranslation on the vacuum. 

We now argue that $Q_T^\text{soft}\ket{0;\psi}$ must be proportional to the insertion of a soft
graviton. We define the supertranslation as a large infinitesimal diffeomorphism. From the diffeomorphism invariance of the gravitational equations of motion we know that $h^T_{\mu\nu} = \mathcal{L}_T
g_{\mu\nu}$ must solve the linearized equations of motion; $h^T_{\mu\nu}$ must be a
graviton. Furthermore, since the supertranslations all commute with $\partial_u$, we know that
$h^T_{\mu\nu}$ must be a zero energy graviton. Let us emphasize that above, following the
literature, we found the BMS diffeomorphisms as the (large) residual gauge symmetry after imposing the Bondi
gauge condition~\eqref{eq:bondi-gauge}, and therefore the $\mathcal{L}_T g_{\mu\nu}$ gives a soft
graviton in Bondi gauge. On the other hand, traditional discussions of gravitational
scattering/quantization are in de Donder gauge, 
\[\label{eq:deDonder}
\nabla^\mu h_{\mu\nu} = \frac{1}{2}\nabla_\nu h.
\] 
These two gauge conditions are inconsistent! This leads to some apparent mismatches in the
discussion of~\cite{Kapec:2015vwa}. In order to explicitly connect Weinberg's soft theorem to
BMS\textsuperscript{0}, one must either discuss the soft theorem in Bondi gauge or find the BMS
transformations in de Donder gauge. In Appendix~\ref{sec:deDond}, we do the latter, where one can
find the leading behavior of the inserted soft graviton explicitly in~\eqref{eq:dDsoftgrav6}.

\section{Strings in Asymptotically Flat Space}
\label{sec:stringasympspc}

We now turn to the discussion of the BMS group and its connection to the soft expansion of string
scattering amplitudes in string theory. For a proper discussion of string theory on asymptotically
flat spacetime in the path integral formalism, a discussion of the path integral measure is
indispensable. We then turn to a discussion of diffeomorphisms in target space and their realization
in the worldsheet theory. In the course of this discussion, we introduce the vertex operator of a
soft graviton $V_{s}(z)$.

As already hinted at in the last subsection, the choice of boundary conditions is extremely
important when talking about large diffeomorphisms and the BMS group in higher dimensions. We will
give a discussion of this from the point of view of BRST cohomology.

We conclude this section with a discussion of Weinberg's soft theorem and how it is realized from
the generalized WT relations introduced in sec. \ref{sec:ward}.

\subsection{The Measure of the Path Integral}

The Polyakov action for a (bosonic) string on a general background manifold with metric $G_{\mu\nu}$ is given by
\[
S_{0} = \frac{1}{2\pi\alpha'}\int d^{2}z\ G_{\mu\nu}(X)\del X^{\mu} \bar{\del}X^{\nu},\label{eq:string}
\]
where we have gauge-fixed the worldsheet metric, and thus there is an additional $bc$ ghost system
$S_{bc}$ added to the above, which appears in the appendix. As mentioned earlier, even though string
theory is a theory of gravity (for the target space), the worldsheet theory is not actually
invariant under spacetime diffeomorphisms. In the worldsheet theory, these correspond to field
redefinitions\footnote{This is not a new point of view~\cite{Evans:1989xq,
    Evans:1989cs,Ovrut:1990ue,Sen:1992pw}.}
\begin{equation}
X^\mu \mapsto X^\mu + \xi^\mu(X).
\end{equation}
This is clearly neither a rigid nor a gauge symmetry of the Polyakov action for general $\xi$. On
the other hand, field redefinitions do not change the physics. This is realized in field theory by
the \emph{equivalence theorem}~\cite{Dyson:1949ha,Kamefuchi:1961sb,Kallosh:1972ap}, which states
that the $\mathcal{S}$-matrix and other observables are invariant under field redefinitions. From the worldsheet
theory point of view, the equivalence between different target space coordinates represents a simple
form of duality between different conformal field theories. However, when using the path integral
formulation of string theory, such a field redefinition may cause the path integral measure to pick
up a Jacobian. Thus it is necessary to correctly define the path integral measure first. In the case
at hand, we know that the spacetime metric will transform under a general (infinitesimal)
diffeomorphism $\xi^\mu(X)$ as
\begin{equation}
\mathcal{L}_\xi G_{\mu\nu} = \xi^\rho\partial_\rho G_{\mu\nu} + G_{\mu\rho}\partial_{\nu}\xi^\rho  + G_{\rho\nu}\partial_\mu\xi^\rho 
   = \nabla_\mu\xi_{\nu} + \nabla_\nu\xi_\mu.
\end{equation}
The correct measure for the path integral of the Polyakov action is the invariant measure (see
e.g.~\cite{Evans:1989cs, Tseytlin:1989si, Tseytlin:1989md}), which can be written schematically as
the infinite product
\[
D^dX \sqrt{-G} = \prod_j d^dX(\sigma_j)\sqrt{-\det G(\sigma_j)}.
\]

To motivate the inclusion of the $\sqrt{-G}$, let us rewrite the Polyakov action in the Hamiltonian
framework. For this discussion, it is convenient to use a Minkowski worldsheet signature,
\begin{equation}\label{eq:string-mink}
S = \frac{1}{4\pi\alpha^\prime}\int d^2\sigma\, G_{\mu\nu}\left(\dot{X}^\mu\dot{X}^\nu - X'^\mu X'^\nu\right),
\end{equation}
so that the canonical momenta are
\begin{equation}
\pi_\mu = \frac{\delta S}{\delta \dot{X}^\mu} = \frac{1}{2\pi\alpha^\prime} G_{\mu\nu}\dot{X}^\nu.
\end{equation}
Thus, expressing the action in terms of phase space variables gives a kinetic term of the form
$G^{\mu\nu}(X)\pi_\mu\pi_\nu$.  The path integral is most correctly defined as the path integral in
phase space of the Hamiltonian action (see eg.~\cite{Ramond:1981pw}):
\[
Z = \int Dq\, D p\, e^{i\int dt\, [p \dot{q} - H(q,p)]}.
\]
Of course, for standard kinetic terms, the Gaussian $Dp$ integral is absorbed into a new measure
$Dq$ and the exponent becomes the action, leaving one with the familiar Feynman path integral. In
cases with nonstandard kinetic terms, such as the worldsheet with target space metric $G_{\mu\nu}$,
the $Dp$ integral can give important contributions. Performing the $D^d\pi$ integral for the string,
brings down a factor of $\sqrt{-\det G}$, the covariant measure. To wit, under an infinitesimal field
redefinition $X^{\mu} \mapsto X^\mu + \xi^{\mu}(X)$, the correct string measure, $D^dX\sqrt{-G}$, is
invariant and therefore a classical discussion should suffice.

\subsection{Target Space Diffeomorphisms}
\label{sec:TargetSpaceDiffeos}

We have established that under diffeomorphisms of the target space, the measure of the path integral
is invariant. In the terminology laid out in sec.~\ref{sec:ward} we understand such field
redefinitions as \emph{general transformations} of the Polyakov action. Under an infinitesimal
change in the field
\[
X^{\mu}(z) \to X^{\mu} + \xi^{\mu}(X(z))
\] 
the action will therefore transform as
\[
\delta S[X] = \int d^{2}z\ V(z) = \frac{1}{2\pi\alpha'}\int d^{2}z\
\left(\nabla_{\mu}\xi_{\nu} + \nabla_{\nu}\xi_{\mu}\right) \del X^{\mu}\bar{\del} X^{\nu} \label{eq:vertop}
\] 
where $\nabla_{\mu}$ is the covariant derivative with respect to the target space metric
$G_{\mu\nu}$. The global part of the target space diffeomorphisms is the group of isometries of the
target space. For these, the generating vector fields $\xi$ satisfy the Killing
equation 
\[
\mathcal{L}_{\xi}G_{\mu\nu}(X) = \nabla_{\mu}\xi_{\nu} + \nabla_{\nu}\xi_{\mu} =0,
\] 
from which we can conclude that Killing vectors do not generate vertex operators, $V(z) = 0$, as it should be. In the first equation we used that the
variation of the action is given by the Lie derivative of the metric. 

The absence of a vertex operator of course implies that there is a
normal WT identity which is consistent with their treatment as symmetries of the action ensuring,
e.g., conservation of energy (for a time-like Killing vector) in target space. More generally,
``small'' diffeomorphisms of the target space with $\mathcal{L}_{\xi}G \neq 0$ will \emph{not} leave
the action invariant.  The resulting vertex operators $V(z)$ correspond to graviton vertex operators with zero-norm light-like polarizations. Any
correlation functions calculated with such insertions will vanish.\footnote{See, e.g.,
  \cite{Polchinski:1998rq} for a discussion. This statement is known as the \emph{cancelled
    propagator} argument.}

The current associated with the infinitesimal diffeomorphism,\footnote{For
  a translation $\xi =a^{\nu}\partial_\nu$ one can compare this to results in~\cite{Polchinski:1998rq}, but
  be mindful of the different normalization of the Noether current.}
\begin{equation}
j_\xi = \frac{1}{2\pi\alpha^\prime}G_{\mu\nu}(X) \xi^\mu(X) \partial X^\nu,\qquad
\tilde{\jmath}_\xi = \frac{1}{2\pi\alpha^\prime}G_{\mu\nu}(X) \xi^\mu(X) \bar{\partial} X^\nu,
\end{equation}
so that the WT identity follows by taking the divergence, $\bar{\partial} j_\xi + \partial
\tilde{\jmath}_\xi$, using the vertex operator $V_\xi$ above, and the contact terms from the
equations of motion. Many authors do not seem to make a sharp distinction between the current (or
its divergence) and the vertex operator. Indeed, up to terms proportional to the equations of motion
the divergence of the current \emph{is} the vertex operator:
\[\label{eq:divJisV}
\bar{\partial} j_\xi + \partial \tilde{\jmath}_\xi = V_\xi 
 + \frac{1}{\pi \alpha^\prime}\xi^\mu G_{\mu\nu}\left(\Box X^\nu + \Gamma^\nu_{\rho\sigma}\partial X^\rho\bar{\partial} X^\sigma\right),
\]
where $\Gamma$ is the target space Christoffel symbol. (Of course, this is just as it should be so
that the contact terms from the Schwinger--Dyson equations give the WT identity.)  The role of the
factors in the WT identity mix with one another depending on the choice of integration domain in
integrating~\eqref{eq:anomWard}. If one integrates over the entire $z$-plane, then by Stokes'
theorem and standard contour-pulling arguments one can drop the contribution of the current in the
WT identity, but must keep the vertex operator and the transformation of the other fields. On the
other hand, one can consider a domain of integration with holes cut out around other insertions,
then the divergence of the current gives the transformation of the fields via Stokes' theorem and
the OPE, and the last term in~\eqref{eq:anomWard} may be dropped. Regardless, the local statement of the
WT identity~\eqref{eq:anomWard} is correct with each term making an independent contribution. We
write it in the above form in order to treat target space isometries in the same manner as general
diffeomorphisms, which necessitates a treatment that includes the usual Noetherian current $j$.

\subsection{Large, Small, and Forbidden Diffeomorphisms on the Worldsheet}
\label{sec:largesmall}

String theory as a theory of gravity computes diffeomorphism-invariant $\mathcal{S}$-matrix elements. How is this realized
on the worldsheet? In fact, the spacetime theory described by the $\mathcal{S}$-matrix elements usually computed
as worldsheet correlators is in de Donder gauge: the graviton vertex operator's momentum must be
transverse to its (traceless) polarization. Moreover, in order for the vertex operator $V_\xi$ to be
$\text{diff}\times\text{Weyl}$ invariant it must satisfy $\Box \xi = 0$: $\xi$ belongs to the
residual gauge symmetry in de Donder gauge. See~\cite{Evans:1989xq} for an early discussion of this
point. If we imagine expanding $\xi$ in Fourier modes, $e^{ik\cdot X}$, then $V_\xi$ takes exactly
the form of a ``pure gauge'' graviton vertex operator. This mode decouples from the theory, as it
must to maintain target space gauge invariance. To show this, one may write $V_\xi$ as the sum of a
total derivative term which may be dropped and a term proportional to $\Box X$, which vanishes due
to the cancelled propagator argument~\cite{Polchinski:1998rq}. On the other hand,
as~\cite{Schulgin:2013xya,Schulgin:2014gra} recently emphasized, at the level of BRST cohomology one
may consider more general diffeomorphisms; the non-Weyl invariant contributions are BRST exact and therefore
continue to decouple.

We would like to classify target space diffeomorphisms given by $\xi^\mu$ as either forbidden,
large, or small from the worldsheet. This discussion closely follows the AdS\textsubscript{3} case
as first laid out in~\cite{Kutasov:1999xu,deBoer:1998pp}. First, let us discuss small
diffeomorphisms. Small diffeomorphisms should result in a trivial WT identity, just as classically
they yield a current that vanishes on-shell.\footnote{This statement follows for any local symmetry
  from ``Noether's second theorem''. Only the ``large'' gauge transformations that make boundary
  contributions give nontrivial currents. See e.g.~\cite{Barnich:2001jy}.} This derives from the
following observations. First the transformation of vertex operators stay in their BRST equivalence
classes from standard arguments, so the last term in~\eqref{eq:anomWard} is BRST trivial. Second, we
can either use~\eqref{eq:divJisV} or the observation that the divergence of the current is BRST
exact with antecedent operator given in~\cite{Schulgin:2013xya, Schulgin:2014gra}
\begin{equation}\begin{gathered}\label{eq:BRST-divj}
s_\xi = c\, G_{\mu\nu}\xi^\mu \partial X^\nu - \tilde{c}\, G_{\mu\nu}\xi^\mu\bar{\partial} X^\nu\\
\{ Q_B,\, s_\xi\} = - 2\pi \alpha^\prime\, c\tilde{c}\,\big(\bar{\partial }j_\xi + \partial j_\xi\big) 
   - c\partial \tilde{c}\, G_{\mu\nu} \xi^\mu \bar{\partial} X^\nu - \tilde{c}\partial c\, G_{\mu\nu}\xi^\mu \partial X^\nu + O(\alpha^\prime)
\end{gathered}\end{equation}
Therefore, the only contribution can come from the equation of motion term. (The last two terms do
not have the right ghost structure to contribute to any correlator.) If $\xi(X)$ is in the form of a
Fourier mode, then one may apply the cancelled propagator argument. Thus the WT identity is a
trivial identity for small diffeomorphisms, as expected.

What breaks down for large, and then forbidden diffeomorphisms? Let us focus on the case when we
integrate the identity~\eqref{eq:anomWard} over the entire worldsheet. Two things can happen. One is
that the cancelled propagator argument can fail: for example, when $\xi(X)$ is constant (i.e.~a
translation), there is no $e^{ik\cdot X}$ term that provides a kinematic region that suppresses
contact terms. The second is more subtle. Equation~\eqref{eq:BRST-divj} continues to hold; however,
the state corresponding to $s_\xi$ is no longer part of the Hilbert space and so one cannot say that
the divergence of the current is BRST exact~\cite{Kutasov:1999xu,deBoer:1998pp}. Scilicet, one gets
nontrivial transformations of the vertex operators and a nontrivial Ward identity. (For
translations, one, of course, gets conservation of target space momentum~\cite{Polchinski:1998rq}.)
The state corresponding to the operator $s_\xi$ fails to be part of the worldsheet Hilbert space
because it is no longer normalizable: the leading term in the $s_\xi s_\xi$ OPE does not fall off
like a power law.\footnote{Remember that the $c\, c$ OPE grows linearly with separation.}

Small diffeomorphisms, then, are given by $\xi(X)$ that are ``good'' (normalizable) operators in the
worldsheet. This includes individual Fourier modes $e^{ik\cdot X}$. Thus, it is useful to consider $\xi$ in
momentum space:\footnote{As we mentioned before,  $\xi$ belongs to the residual gauge symmetry in de Donder gauge. Therefore the Fourier transform should be taken over momenta satisfying $k^{2}=0$.}
\[
\xi^\mu(X) = \int \frac{d^dk}{(2\pi)^d} \xi^\mu(k) e^{ik\cdot X}.
\]
Then perform an infinitesimal diffeomorphism $X^\mu \mapsto X^\mu + \xi^\mu(X)$ on a
correlator. For simplicity, consider a tachyon two-point function,
\begin{equation}
\braket{\tilde{c}c\, e^{i k_1\cdot X}(z_1)\, \tilde{c} c e^{ik_2\cdot X}(z_2)} \propto \delta(k_1 + k_2).
\end{equation}
Performing the diffeomorphism adds an extra term of the form
\begin{equation}
i  (k_1 + k_2)_\mu\xi^\mu\big(k_1+k_2\big),
\end{equation}
which vanishes when one imposes momentum conservation if $\xi^\mu(k)$ is regular at $k=0$. Recall,
that this is being added to a $\delta$-function. In order for $\xi$ to give a physical shift, it
must be at least as IR divergent as the leading term. This means in momentum space $\xi^\mu(k)$
should be at least as singular as $\delta(k)$ near $k=0$. In position space, this means that any
diffeomorphism that falls off faster than translations near the boundary is ``small''. Anything that
grows at least as fast as translations is ``large''. 

As for forbidden diffeomorphisms, one can consider $\xi(X)$ for which either the vertex operator
$V_\xi$ or the transformation of other vertex operators $\delta_\xi V$ is not a good operator in the
worldsheet theory: a diffeomorphism $\xi$ is forbidden if $\delta_\xi G_{\mu\nu} = \mathcal{L}_\xi
G_{\mu\nu}$ or the transformation $\delta_\xi V$ not normalizable. To summarize, the large
diffeomorphisms live in the sweet spot in which there is a nontrivial, but well defined WT identity
on the worldsheet. 

\subsection{Boundary Conditions in $d$-Dimensions}
\label{sec:bdycond}

Let us consider these conditions for a string in Minkowski space. The condition that $s_\xi$ be
normalizable is suggestive of a condition that the components $\xi^\mu(X)$ be square-integrable
(elements of $L^2(\mathbb{R}^d)$). Applying this criterion to $\delta_\xi G_{\mu\nu}$ would
essentially impose the stronger boundary conditions in~\eqref{eq:radiative-BCs}, which would
eliminate any symmetry enhancement. With some thought, one should realize that this is too strong a
condition. This would suggest that all the usual vertex operators one considers in string theory are
not ``good'' operators: one would be required to only consider square-integrable wave
packets. Moreover, it would seem to suggest that an operator could be made non-normalizable just by
adding an extra field.  Instead, we allow constant behavior in some directions; the exception being
when it is constant in all directions, which we already discussed above. This suggests the
asymptotic symmetry algebra is defined in terms of equivalence classes
\[
\xi^\mu \simeq \xi^\mu + o\left(r^0\right),
\]
where the ``little-o'' notation indicates strictly subleading behavior, and furthermore that we only allow
diffeomorphisms such that
\[
\mathcal{L}_\xi G_{\mu\nu} \simeq o\left(r^0\right).
\]
In addition to the above condition, we impose an analyticity condition: we demand that $\delta_\xi
G_{\mu\nu}$ at leading order does not introduce a branch cut. Both from the worldsheet and the
spacetime points of view, a branch cut in $r$ is somewhat problematic. 

By flushing out the above conditions, one may deduce the boundary conditions that string theory
implies for asymptotically flat spacetime in $d$ non-compact dimensions. Starting from flat space,
acting with the space of allowed diffeomorphisms gives the space of allowed asymptotic metrics, and
therefore fall-off conditions on the asymptotically flat metric. A few caveats are in
order. First, our discussion is entirely on-shell, which means we cannot see any curvature
conditions of the kind suggested in~\cite{Kapec:2015vwa}. Moreover, our analysis is only in flat
space, and not in a general asymptotically flat space. Nonetheless, what we do have is sufficient to
determine the asymptotic symmetry algebra, and ultimately get a soft theorem.

In particular, the string worldsheet suggests the following boundary conditions in Min\-kow\-ski space:
\begin{equation}
g_{\mu\nu} = \eta_{\mu\nu} + O\left(\tfrac{1}{r}\right).
\end{equation}
Transforming this condition into Bondi coordinates is straightforward and yields
\begin{equation}\begin{gathered}\label{eq:string-BCs}
\delta g_{uu}\sim \delta g_{ur} \sim \delta g_{rr} \sim O\left(\tfrac{1}{r}\right)\\
\delta g_{uA} \sim \delta g_{rA} \sim O\left(r^0\right)\\
\delta g_{AB} \sim O\left(r^1\right),
\end{gathered}\end{equation}
with small diffeomorphisms defined by
\begin{equation}
\xi_\text{small} = o\left(r^0\right)\partial_u + o\left(r^0\right)\partial_r + o\left(r^{-1}\right) \partial_A.
\end{equation}
Comparing with the boundary conditions in~\eqref{eq:asymp-flat} is worthwhile. The first thing to
note is that since we are not in Bondi gauge, $\delta g_{rr}$ and $\delta g_{rA}$ are not forced to
vanish and there is no determinant condition. The second thing to note is the different fall-off
condition on $g_{ur}$. 

Since spacetime indices are internal labels on the worldsheet, any boundary conditions that one can
derive from string theory in this manner must have the same structure as above. Starting with a
``scalar'' condition in Minkowski coordinates (an isotropic condition) makes it clear that one
should have complete freedom in choosing coordinates on the sphere $S^{d-2}$. This suggests an
additional freedom similar to the discussion in~\cite{Campiglia:2014yka}.  We return to this point
in the conclusion.

One can and should ask how rigid the above boundary conditions are, since in the spacetime theory
there seemed a fair amount of freedom. The authors do not know of any analysis of this interesting
question, and whether it is sensible to imagine, for example, adding a boundary condition in the
field space of the worldsheet path integral. This is especially of interest given alternate
quantizations in AdS/CFT. Regardless of their uniqueness, the above analysis suggests that the
boundary conditions~\eqref{eq:string-BCs} are consistent within string theory.

\subsection{The Soft Theorem, Part II: String Theory}
\label{sec:softSec}

We now turn to a discussion of the soft theorem from BMS supertranslations as manifested in string
theory. We concentrate on the supertranslations, so throughout this section, the function
$Y^{A} = 0$, while the function $T$ is kept arbitrary. The ``large'' part of these transformations
is 
\begin{align}\label{eq:largeST}
\xi^{u} &= T(\theta) \notag\\
\xi^{r} &= \frac{\Delta T}{d-2} + \frac{u}{(d-4)(d-2)r}\Delta (\Delta +d-2) T\\
\xi^{A} &=-\frac{1}{r} D^A\left[ T + \frac{u}{(d-2)r}\left(\Delta  + d-2\right)T\right]\notag
\end{align} 
in Bondi coordinates and de Donder gauge. A derivation of these components may be found in apdx.~\ref{sec:deDond}. The transformation of the action is given in \eqref{eq:vertop} for general
target space diffeomorphisms. Using the vector field \eqref{eq:largeST} the leading terms in the Lie
derivative of the metric are the $G_{uA}$ and $G_{AB}$ components. The vertex operator resulting
from such a transformation is then
\begin{multline}\label{eq:vertopso}
(d-2)\pi\alpha' V_{s}(\sigma) = -D_{A}\left(\Delta + d-2\right)T(\theta)\left(\del u \bar{\del} \theta^{A} 
+ \bar{\del} u \del \theta^{A}\right)\\
+ r \left(\gamma_{AB}\Delta - (d-2) D_{A}D_{B}\right) T(\theta)\ \del \theta^{A}\bar{\del}\theta^{B}
\end{multline}
For $T = c$, a constant, the vertex operator vanishes. As constant $T$ corresponds to a time translation, this is a good validity check. The $d-1$ spacial translations, for which $T$ satisfies \[(\Delta +d-2) T = 0\] also lead to a vanishing vertex operator. The resulting unbroken WT identities are then the usual statement of conservation of momentum. 

As we explained above, since $V_{s}$ is generated by a target space diffeomorphism, it inserts a graviton into the theory. Contrary to the case of small diffeomorphisms, however, the vertex operators of large diffeomorphisms correspond to insertions of physical, \emph{zero momentum} gravitons and the canceled propagator argument does not hold. Note that $V_{s}$ appears at the same order in the $r$ expansion, its form is only altered by the specifics of the Laplacian, covariant derivative and dimension of the sphere $S^{d-2}$. In particular it does not depend on the fall-off behavior of gravitational radiation.

Consider now a correlation function $\langle V_{1}\cdots V_{n}\rangle$ where $V_{i}$ are bosonic
string vertex operators\footnote{A similar statement should hold for the superstring, but the form
  of the vertex operator \eqref{eq:vertopso} will be altered by fermionic contributions.}. We
 use the generalized WT identity \eqref{eq:anomWard} in its integrated form. After reinstating
the integral over the worldsheet, the total differential of the current $j$ can be written as a
contour integral. This form allows us to use a standard contour pulling argument to show that the
left hand side of the identity must vanish. When the contour encircles all vertex operators, we may
shrink it to little circles around the insertions and the insertion point of $j$ to pick up the
residues at the single poles in the OPE of $j^{a}$ and the $V_{i}$. Since the current is not
actually conserved, an additional term appears, which is of course the insertion of
$V_{s}(z)$. Alternatively, we may understand the contour as encircling an empty patch on the
worldsheet. In this case, the integral must vanish and so the left hand side of the generalized WT
identity must vanish.  This is true even though a new (local) operator appears on the right hand
side for the case of a general transformation. The only condition for this argument to work is that
the transformation is well behaved on the empty patch on the worldsheet (see
sec.~\ref{sec:largesmall}).

We may now turn to the second term on the right hand side of the generalized WT identity. In flat
coordinates $X^{\mu}$ the vertex operators assume the usual form 
\[
V_{i}(z,\bar{z}) =\zeta_{\mu\nu}\del X^{\mu}\bar{\del} X^{\nu}\exp(i k_{i}.X)
\] 
for gravitons. The polarization tensor $\zeta_{\mu\nu}$ is naturally given in TT
gauge. Because of this, general variations of such vertex operators can be written as operators
acting on the vertex operator itself \[\delta V_{i} =
\left(\frac12(\del_{\mu}\xi_{\nu}-\del_{\nu}\xi_{\mu})S_{i}^{\mu\nu} + i k_{i}.\xi \right)V_{i} +
{\rm small}\label{eq:varVert}.\] $S_{i}^{\mu\nu}$ is the spin dependent representation of the
Lorentz algebra, or the spin angular momentum operator. We shall refer to the first part in
\eqref{eq:varVert} as the \emph{spin part} and the second part as the \emph{orbital part} of the
variation. The spin part is small for supertranslations, such that
we can ignore it here. Superrotations have terms superleading over the supertranslation part. For these, the spin part cannot be neglected and they provides the spin dependent part of
the subleading soft theorem \cite{Cachazo:2014fwa}.

The orbital part is universal for any vertex operator, i.e., every particle in the (bosonic) string
spectrum has this particular piece. Using~\eqref{eq:largeST}, restricting to supertranslations,
and discarding any small terms we find
\[
\delta_{\xi} V_{i} = i E_{i} T(\theta_{i}) V_{i}.
\]
The function $T$ is a function of the angular variables at the point of insertion of $V$ and can be
chosen freely. The most natural choice in the context of the soft theorem would be to set it to a
$\delta$-distribution aligning the null modes of the angular worldsheet variables $\theta_{i}$ with
the angular variables on the boundary $\theta_{0}$ as is done in the literature. This would
correspond to the insertion of a single soft graviton. But we are not forced to do so.  The
aforementioned parametrization of the soft factor is always possible, even in higher dimensions
\cite{Kapec:2015vwa}. Thus we find that the variation of the string vertex operators corresponds to
the ``hard part'' of the action of the supertranslation charge. It follows that
\[
\int d^{2}z \langle V_{s}(z)V_{1}\cdots V_{n}\rangle = \sum_{i=1}^{n} E_{i}T(\theta_{i})\langle V_{1}\cdots V_{n}\rangle.
\] 

Take note of the fact that this statement does not depend on $\alpha'$ apart from the overall factor
appearing in \eqref{eq:vertop}.\footnote{Of course this $\alpha'$ in $V_{s}$ gets canceled against
  an $\alpha'$ from the OPE. } This is as expected from direct calculations performed
in~\cite{Schwab:2014sla,Schwab:2014fia,Bianchi:2014gla} and from the fact that this is a low-energy
statement. While these results were derived in a bosonic string setting there is nothing that
indicates a different result for the superstring and indeed the result should be the same.

\section{Conclusion and Outlook}

While the extended soft theorems in gauge theory and gravity have spawned an extensive literature
recently, the fate of BMS symmetry in string theory or in higher dimensional spacetimes had not been
adequately addressed so far. In this paper we made an attempt to fill in this gap by presenting a
detailed investigation of large diffeomorphisms of asymptotically flat target spaces from the point
of view of bosonic string theory. We have shown that these imply the existence of soft graviton
modes and Weinberg's soft theorem also in string theory, but the correlator-like structure of the
$\mathcal{S}$-matrix of first quantized string theory makes the derivations in the two theories
rather different. Let us note that while we haven't discussed this topic in any detail here, an
extension of this result to superstring theory should be essentially effortless. The behavior of the
Kalb-Ramond field $B_{\mu\nu}$ (for the corresponding soft theorem, see
\cite{DiVecchia:2015oba,Bianchi:2015yta}) under BMS transformations can be investigated by keeping
the holomorphic and antiholomorphic current independent of each other.

We would like to point out an important detail about string theory here. String theory is effectively formulated in de Donder gauge. Bondi gauge and de Donder gauge are incompatible gauges in the sense that one cannot have the de Donder gauge condition as well as the Bondi requirement that $g_{rr} = g_{rA} = 0$. For this reason, it was important to rederive the BMS transformations in de Donder gauge using $\Box \xi = 0$ subject to the correct boundary conditions.

One result which we find particularly intriguing are the boundary conditions in $d$ dimensions found
in sec.~\ref{sec:bdycond}. First of all, these conditions do not depend on the number of dimensions
or the distinction between even and odd $d$. This is particularly appealing when considering the
connection between Weinberg's soft theorem and BMS. As Weinberg's soft theorem is essentially
independent of the number of spacetime dimensions, so should be the BMS symmetry which generates
it. In particular, the dimension of the sphere should not be important. Secondly, and more
importantly, it is intriguing to think that string theory on asymptotically flat spacetimes might
not just prefer but actually requires a certain set of boundary conditions. Such a statement is
rather different from the AdS case where there are alternative quantizations depending on the choice
of boundary conditions.

The question remains how this program might be extended to superrotations. And, indeed, how to
define superrotations in higher dimensions.  As for the latter question, it is clear that a
structure like a Virasoro algebra is ruled out in higher dimensions. To find a Virasoro algebra,
however, might not be necessary at all. In~\cite{Campiglia:2014yka}, \citeauthor{Campiglia:2014yka}
show that a generalization of the extended BMS group in four dimensions still yields the correct
soft theorem. The authors assumed that superrotations can be generalized to general smooth vector
fields on the sphere, i.e., the assumption that these vector fields need to be conformal Killing
vector fields is dropped. The smooth vector fields on the sphere form the group ${\rm Diff}(S^{2})$,
such that the extended group of asymptotic symmetries becomes \[G = T \rtimes {\rm Diff}(S^{2}).\]
The apparent downside to this approach is that one needs to allow for a ``dynamic'' metric on the
sphere. We have commented on this in the text: it doesn't seem problematic to allow for a choice of
coordinates on the sphere. Also, compared to the Virasoro algebra of the extended BMS group in four
dimensions, there seems to exist no obvious obstruction to generalizing the group of diffeomorphisms
of the sphere to higher dimensions. Note, however, that superrotation symmetry appears to be broken
at the first loop level \cite{Bern:2014oka} in four dimensions. This should translate into an issue
with superrotations and the path integral measure in the case of higher genus worldsheet
topologies. We will leave this problem for another time.

\paragraph{Acknowledgements}

The authors would like to thank Antal Jevicki and Marcus Spradlin for their support and helpful
comments during the preparation of this manuscript. We also would like to thank Robert de Mello Koch
and Chung-I Tan for helpful discussions. SA is grateful for discussions with Michael Zlotnikov on a
related project, and correspondence with Sujay Ashok. This work is supported by the US Department of
Energy under contract DE-FG02-11ER41742.

\appendix
\addcontentsline{toc}{section}{Appendices}
\addtocontents{toc}{\protect\setcounter{tocdepth}{-1}}

\section{The Ghost Sector}
\label{sec:conventions}

Here we briefly review the BRST quantization of the string following~\cite{Polchinski:1998rq}. The
covariant quantization of the string via Faddeev--Popov adds a ghost action to give the correct
measure for the gauge-fixed string. The ghost action takes the form
\begin{equation}
S_\text{gh} = \frac{1}{2\pi}\int d^2 z\, (b\bar{\partial} c + \tilde{b}\partial \tilde{c}).
\end{equation}
So that the total action (after gauge-fixing the worldsheet metric) is $S_0$ from
Equation~\eqref{eq:string} plus $S_\text{ghost}$. The sum, $S_0 + S_\text{gh}$, enjoys a
reincarnation of the original $\text{diff}\times \text{Weyl}$ symmetry as the Grassmann-odd BRST symmetry:
\begin{equation}\begin{aligned}
\delta_B X^\mu &= i\epsilon(c\partial X^\mu + \tilde{c}\bar{\partial} X^\mu)\\
\delta_B b &= i\epsilon (T^X + T^\text{gh}) &
\delta_B \tilde{b} &= i\epsilon(\tilde{T}^X + \tilde{T}^\text{gh})\\
\delta_B c &= i\epsilon c\partial c &
\delta_B \tilde{c} &= i\epsilon\tilde{c}\bar{\partial}\tilde{c},\\
T^X &= -\frac{1}{\alpha^\prime}G_{\mu\nu}(X)\partial X^\mu \partial X^\nu &
\tilde{T}^X &= -\frac{1}{\alpha^\prime}G_{\mu\nu}(X)\bar{\partial} X^\mu \bar{\partial}X^\nu \\
T^\text{gh} &= (\partial b) c - 2 \partial(bc) &
T^\text{gh} &= (\bar{\partial}\tilde{b})\tilde{c} - 2 \bar{\partial}(\tilde{b}\tilde{c}).
\end{aligned}\end{equation}
The BRST transformation is nilpotent with the physical states given by the BRST cohomology: the
physical Hilbert space is the space of closed states moded out by the space of exact states.

\section{BMS in de Donder}
\label{sec:deDond}

There are two gauges that appear in the discussion: Bondi~\eqref{eq:bondi-gauge} and de
Donder~\eqref{eq:deDonder}. The latter is the linearization of harmonic gauge $\Box x^\mu = 0$,
whereas~\eqref{eq:bondi-gauge} is already a condition on the nonlinear theory. If one notices that
the Bondi coordinates for flat space are not harmonic, then one should already be suspicious;
however, we are interested in imposing conditions on the linearized theory living in Bondi
coordinates.

A first attempt to address the question would be to start with a standard TT gauge (de
Donder with $h_{\mu\nu}U^\nu = {h_\mu}^\mu = 0$ for a constant timelike vector $U^\nu$) graviton in
Minkowski coordinates and transform into Bondi coordinates. In which case, one finds that
conditions~\eqref{eq:bondi-gauge} set $h_{\mu\nu} \equiv 0$. This is perhaps not surprising since
one is imposing too many gauge conditions. Instead one may wish to replace the TT gauge condition
with the Bondi condition; they even look similar. A hint that this might not be a good idea is that
the BMS vector fields~\eqref{eq:BMSgen} have nonvanishing d'Alembertian, and thus are not residual
gauge transformations of de Donder. The de Donder condition (after imposing Bondi)
implies
\begin{equation}
h_{rr} = 0\qquad
D^A h_{uA} = r^2\partial_rh_{uu} + (d-2) r h_{uu}\qquad
D^B h_{AB} = r^2 \partial_rh_{uA} + (d-2) r h_{uA}.
\end{equation}
The $uu$ and $ur$ components of the Einstein tensor force $h_{uu} = 0$. (One uses the fact that the
Laplacian on the sphere is negative definite.) Similarly, the $uA$ and $rA$ components force $h_{uA}
= 0$. Then one is left with $h_{AB}$ with the trace condition $\gamma^{AB}h_{AB}$ coming from the
linearization of the Bondi determinant condition. This has the right number of degrees of freedom,
but then one imposes $D^Bh_{AB} = 0$, which removes physical degrees of freedom.

Since de Donder is usually the preferred gauge condition for discussing radiation and quantization,
it is useful to find the supertranslation generators in de Donder instead of Bondi gauge. As far as
the authors know, this has not been presented in the literature. To state the problem, we are
looking for harmonic vector fields $\Box \xi^\mu = 0$, which respect a de Donder version of the
boundary conditions~\eqref{eq:asymp-flat}. We want the solution in flat space with Bondi
coordinates. We take a slight shortcut by looking for harmonic vector fields with the same leading
behavior as~\eqref{eq:BMSgen}, since the leading piece should be ``large'' and gauge invariant.

In Bondi coordinates, for a vector
\begin{equation}
\xi = F\partial_u + R\partial_r + \frac{1}{r} S^A\partial_A,
\end{equation}
Laplace's equation takes the form
\begin{equation}\begin{aligned}
\Box \xi^u &= \Box_S F + \frac{1}{r^2}\left((d-2)R + 2 \,D_AS^A\right)\\
\Box \xi^r &= \Box_S R - \frac{1}{r^2}\left((d-2)R + 2 \,D_AS^A\right)\\
r \Box \xi^A &= \Box_S S^A + \frac{1}{r^2}(2D^A R - S^A),
\end{aligned}\end{equation}
where
\begin{equation}\label{eq:boxS}
\Box_S f = f'' - 2 \dot{f}' + \frac{d-2}{r} (f' - \dot{f}) + \frac{\Delta}{r^2}f,
\end{equation}
with primes and dots denoting $r$- and $u$-derivatives, respectively. We are using $\Delta$ to
denote the Laplacian on the unit sphere.  Rather than Fourier transforming $u$ as might seem
natural, let us look for solutions that are polynomial in $u$. In fact, let us make an ansatz 
that $F$, $R$, and $S^A$ are functions of $x=\frac{u}{r}$ and $\theta^A$
only. Then, the above equations become
\begin{equation}\begin{aligned}
r^2 \Box \xi^u &= x(x+2) F'' - (d-4)(1+x)F' + \Delta_\Omega F + (d-2) R + 2 D_A S^A = 0\\
r^2\Box \xi^r &= x(x+2)R'' - (d-4)(1+x)R' + \Delta_\Omega R - (d-2) R - 2 D_A S^A = 0\\
r^3\Box \xi^A &= x(x+2){S''}^A - (d-4)(1+x){S'}^A + \Delta_\Omega S^A + 2 D^A R - S^A = 0
\end{aligned}\end{equation}
We are interested in the behavior for large $r$ and fixed $u$, which corresponds to $x\to 0$. The
point $x=0$ is a regular singular point of the above coupled DEs. To find a series solution let us
look at the indicial equation by trying a power law
\[
F \simeq F_0 x^\lambda\qquad
R \simeq R_0 x^\lambda\qquad
S^A \simeq S^A_0 x^\lambda.
\]
We see that the leading behavior of the three equations decouple and give the condition
\begin{equation}
2 \lambda(\lambda - 1) - (d-4)\lambda = 0\qquad \Longrightarrow\qquad \lambda = 0 \text{ or } \lambda =\frac{d-2}{2}.
\end{equation}
We are interested only in the $\lambda = 0$ solution. Therefore, let us plug in with
\begin{equation}
F(x, \theta) = \sum_{n=0}^\infty F_n(\theta) x^n,
\end{equation}
and likewise for $R$ and $S^A$ into the equation for $\Box\xi^\mu = 0$. Matching powers of $x$, one
finds the following recursion relations:
\begin{equation}\begin{aligned}
2(n+1)\big(n-\tfrac{d-4}{2}\big)\,F_{n+1} &= -\left[\big(n(n-d+3)+\Delta\big)F_n + (d-2) R_n + 2 D_A S^A_n\right]\\
2(n+1)\big(n-\tfrac{d-4}{2}\big)\,R_{n+1} &= -\left[\big(n(n-d+3)+\Delta\big)R_n - (d-2) R_n - 2 D_A S^A_n\right]\\
2(n+1)\big(n-\tfrac{d-4}{2}\big)\,S^A_{n+1} &= -\left[\big(n(n-d+3)+\Delta\big)S^A_n +2D^A R_n- S^A_n\right].
\end{aligned}\end{equation}
Note that the relations degenerate in even dimensions when $n=\frac{d-4}{2}$, which corresponds to
the second asymptotic behavior, $x^\frac{d-2}{2}$. Fortunately, we are interested in the behavior
exactly before the above relations falter. (The singular equation for the coefficient of
$x^\frac{d-2}{2}$ is cured by adding $\log x$ times the other solution.) One can also find exact
solutions in terms of Legendre functions by decomposing into spherical harmonics, but we find the
series solution more convenient.

We are interested in solutions for the following boundary conditions
\begin{equation}
F_0 = T(\theta) \qquad
R_0 = \frac{\Delta T(\theta)}{d-2}\qquad
S^A_0 = - D^A T(\theta).
\end{equation}
Since everything will be written in terms of the scalar function $T(\theta)$, let's put in as an ansatz
\begin{equation}
S^A = D^A\Phi,
\end{equation}
and use the commutator on the sphere,
\begin{equation}
\Delta(D^A \Phi) = D^A(\Delta + d-3)\Phi,
\end{equation}
to find
\begin{equation}\begin{aligned}
2(n+1)\big(n-\tfrac{d-4}{2}\big)\,F_{n+1} &= -\left[\big(n(n-d+3)+\Delta\big)F_n + (d-2) R_n + 2 \Delta \Phi_n\right]\\
2(n+1)\big(n-\tfrac{d-4}{2}\big)\,R_{n+1} &= -\left[\big(n(n-d+3)+\Delta\big)R_n - (d-2) R_n - 2 \Delta \Phi_n\right]\\
2(n+1)\big(n-\tfrac{d-4}{2}\big)\,\Phi_{n+1} &= -\left[\big(n(n-d+3)+\Delta\big)\Phi_n +2 R_n + (d-4)\Phi_n\right].
\end{aligned}\end{equation}

To the first subleading correction, then
\begin{multline}
\xi = T(\theta)\partial_u + \left[\frac{\Delta T}{d-2} + \frac{u}{(d-4)(d-2)r}\Delta (\Delta +d-2) T\right]\partial_r \\
  -\frac{1}{r} D^A\left[ T + \frac{u}{(d-2)r}\left(\Delta  + d-2\right)T\right]\partial_A + \text{subleading},
\end{multline}
which gives a metric perturbation of the form
\begin{multline}\label{eq:dDsoftgrav6}
h = -\frac{2}{(d-4)(d-2)r}\Delta(\Delta +d-2)T du^2 + \frac{2u}{(d-4)(d-2)r^2}\Delta(\Delta+d-2)Tdudr\\
   -\frac{2}{d-2}D_A\left(2+\frac{u}{(d-4)r}\Delta\right)(\Delta + d-2)Tdud\theta^A
   +\frac{4u}{(d-2)r}D_A(\Delta + d-2)T drd\theta^A\\
  +\left[2r\left(\gamma_{AB}\frac{\Delta}{d-2}-D_AD_B\right) + \frac{2u}{d-2}\big(\gamma_{AB}\frac{\Delta}{d-4}- D_AD_B\big)(\Delta + d-2) \right]Td\theta^Ad\theta^B.
\end{multline}
Note that in six dimenions $h_{AB}$ contains terms both of order $r$ and order $r^0$. The leading
piece is a new soft degree of freedom that arises from relaxing the boundary conditions (which the
authors like to refer to as the ``stranslaton''); whereas, the subleading piece is where the
(propagating) radiative data enters in six dimensions. For general $d$ the series solution shifts the radiative data.

\printbibliography
\end{document}